\documentclass[%
 aip,
 amsmath,amssymb,
preprint,%
]{revtex4-1}

\usepackage{graphicx}% Include figure files
\usepackage{dcolumn}% Align table columns on decimal point
\usepackage{bm}% bold math
\usepackage[utf8]{inputenc}
\usepackage[T1]{fontenc}
\usepackage{mathptmx}
\usepackage{etoolbox}
\usepackage[dvipsnames]{xcolor}
\usepackage{epstopdf}

\makeatletter
\def\@email#1#2{%
 \endgroup
 \patchcmd{\titleblock@produce}
  {\frontmatter@RRAPformat}
  {\frontmatter@RRAPformat{\produce@RRAP{*#1\href{mailto:#2}{#2}}}\frontmatter@RRAPformat}
  {}{}
}%
\makeatother
\begin{document}

\preprint{AIP/123-QED}

\title{Octave-Spanning Terahertz Quarter-Wave Plates Based on Over-Coupled Fabry-P\'{e}rot Resonances in Reflective Metal-Dielectric-Metal Metasurfaces}
\author{Tae Gwan Park}
\author{Chun-Chieh Chang}
\author{Antoinette J. Taylor}
\author{Abul K. Azad}
\author{Hou-Tong Chen}
 \email{chenht@lanl.gov.}
 \affiliation{Center for Integrated Nanotechnologies, Los Alamos National Laboratory, Los Alamos, NM 87545, USA}
 \date{\today}% It is always \today, today, but any date may be explicitly specified

\begin{abstract}
Compact devices for broadband polarization control in the terahertz (THz) regime are challenging due to the intrinsic phase dispersion of birefringent materials and resonant structures. Here, we demonstrate high-performance broadband THz quarter-wave plates based on over-coupled metal-dielectric-metal reflective metasurfaces. The devices operate as single-port anisotropic Fabry-P\'{e}rot cavities in which the phase dispersion of over-coupled resonances is engineered to produce an approximately constant relative phase delay between orthogonal field components. By tailoring the metasurface geometry, efficient linear-to-circular polarization conversion is achieved while maintaining high reflectance. Four complementary metasurface designs, operated at an incidence angle of $45^\circ$, collectively cover the 0.25--3 THz frequency range accessible to a typical THz time-domain spectroscopy system. Each device exhibits an approximately octave-wide bandwidth with an axial ratio below 3\,dB and polarization conversion efficiencies exceeding 80\% across most of the operating band. Systematic optimization suppresses coupling to higher-order diffraction and surface wave modes, further extending the usable bandwidth while preserving the required phase relationship. The metasurfaces are compatible with wafer-scale fabrication, and experimental results show excellent agreement with simulations. These findings establish over-coupled reflective metasurfaces as a robust and versatile platform for broadband THz polarization control.
\end{abstract}

\maketitle

\section{\label{sec:introduction}Introduction}
Circularly polarized terahertz (THz) radiation is indispensable for a wide range of applications, including chiral spectroscopy and molecular identification \cite{Xu_Allen_2003_Astrobio,Damari_2016_PRL,Choi_2019_NatMater,Choi_2022_JACS}, all-optical manipulation of magnetization without external magnetic fields \cite{Stanciu_2007_PRL,Lambert_2014_Science,Mangin_2014_NatMater}, and the selective probing or excitation of spin waves (magnons) \cite{Kampfrath_2011_NatPhoton,Luo_2025_NatCommun} and chiral phonons \cite{Baydin_2022_PRL,Tauchert_2022_Nature,Luo_2023_Science}. These applications motivate the development of high-performance linear-to-circular polarization converters (i.e., quarter-wave plates) in the THz regime, particularly because most coherent THz sources emit linearly polarized radiation. Conventional narrowband quarter-wave plates can be realized using low-loss birefringent materials such as quartz \cite{Grischkowsky_1990_JOSAB,Castro-Camus_2007_OE} and liquid crystals \cite{Hsieh_2006_OL}, as well as artificially engineered anisotropic effective media exhibiting form birefringence \cite{Flanders_1983_APL,Richter_1995_AO}, for example, subwavelength silicon gratings \cite{Kadlec_2008_OL,Saha_2010_IEEE}. A strategy widely used in the optical regime and extendable to broadband THz frequencies involves stacking multiple plates with carefully selected thicknesses and relative orientations to compensate for wavelength-dependent phase retardation, thereby forming achromatic THz wave plates \cite{Masson_2006_OE,Chen_2013_OptCommun,Wu_2020_JIMTW}. However, because of the relatively small birefringence of available materials and the intrinsically long THz wavelengths, such devices are typically bulky and often require thicknesses on the order of several centimeters. Prism-based polarization elements, such as Fresnel rhombs \cite{Shan_Heinz_2009_OE,Kawada_2014_OL}, can provide truly achromatic THz quarter-wave retardation by generating polarization phase delay through total internal reflection rather than material birefringence. Nevertheless, these devices remain physically large and are not readily compatible with compact or integrated THz systems. 

The advent of metamaterials and metasurfaces has significantly advanced the control of electromagnetic waves, providing versatile platforms for manipulating amplitude, phase, and polarization with subwavelength spatial resolution in ultrathin planar architectures \cite{Chen_2016_ROPP}. In single-layer metasurfaces composed of arrays of anisotropic resonant subwavelength elements, the reflection and/or transmission responses exhibit distinct resonant frequencies for incident waves polarized along two orthogonal directions. Consequently, the corresponding amplitude and phase spectra exhibit different resonant dispersions, enabling quarter-wave retardation at a target frequency located between the two resonances while preserving equal output amplitudes \cite{Strikwerda_Averitt_2009_OE,Peralta_OHara_2009_OE}. However, similar to conventional single-plate birefringent quarter-wave plates, such metasurface designs typically operate over a relatively narrow bandwidth. This constraint can be mitigated using metasurfaces based on the geometric Pancharatnam-Berry (PB) phase \cite{Bomzon_Hasman_2002_OL,Chen_Zentgraf_2012_NatCommun,Lin_Brongersma_2014_Science}. With suitable designs, the left-handed and right-handed circular polarization components of an incident linearly polarized wave can be spatially separated, thereby producing circularly polarized waves of opposite handedness. Nevertheless, for broadband THz radiation with an extremely large fractional bandwidth, the generated circular polarization exhibits frequency-dependent propagation directions \cite{Jia_Zhou_2019_LSA} or focal lengths \cite{Wang_2015_OE}. Such frequency-dependent spatial dispersion poses a significant challenge for applications in which THz time-domain spectroscopy (THz-TDS) is employed to measure THz pulses over a wide spectral range.

Introducing additional layers and exploiting interlayer coupling can lead to electromagnetic functionalities and performance levels that are unattainable with single-layer metasurfaces. Such multilayer metasurface architectures have facilitated a wide range of advanced phenomena and functionalities, including perfect absorption \cite{Landy_Padilla_2008_PRL,Tao_2008_PRB}, antireflection \cite{Chen_2010_PRL_ARC,Huang_Chen_2017_ACSPhoton}, Huygens' surfaces \cite{Pfeiffer_Grbic_2013_PRL}, asymmetric transmission \cite{Pfeiffer_Grbic_2014_PRL}, and advanced polarization control \cite{Hao_Zhou_2007_PRL,Grady_Chen_2013_Science,Cong_Zhang_2013_APL,Cong_Zhang_2014_LPR,Lee_Withayachumnankul_2018_OE,Chang_2019_PRL,You_2020_APLPhoton,Li_2024_ChinPhysLett}. In the THz regime, broadband transmissive quarter-wave plates have been demonstrated using multilayer metasurfaces. For example, a two-layer metasurface composed of twisted metallic gratings achieved a bandwidth of 0.4\,THz (from 1.0 to 1.4\,THz and from 1.4 to 1.8\,THz), although accompanied by noticeable insertion loss \cite{Cong_Zhang_2014_LPR}. Subsequently, a three-layer transmissive THz metasurface demonstrated broadband quarter-wave retardation with a fractional bandwidth of about 53\% \cite{You_2020_APLPhoton}. Operating in reflection at an incidence angle of $45^\circ$, two-layer metasurfaces have further enabled ultra-broadband THz linear-to-circular polarization conversion, achieving more than one octave of bandwidth (fractional bandwidth \textasciitilde80\%) with near-unity conversion efficiency \cite{Chang_2019_PRL}. Devices based on this architecture have also been utilized to generate broadband, circularly-polarized, strong-field THz radiation \cite{Li_2024_ChinPhysLett}.

In this work, we present a set of reflective metal-dielectric-metal metasurfaces that collectively span the entire THz frequency range accessible to a typical THz time-domain spectroscopy system (0.2--3\,THz). Broadband spectral coverage is achieved using four complementary devices, each designed to operate within a specific subband of this range. The metal-dielectric-metal metasurfaces function as single-port cavity resonators in which reflective Fabry-P\'{e}rot resonances operate in the over-coupled regime. By carefully tailoring the geometric parameters of the anisotropic subwavelength elements and the thickness of the dielectric spacer, we achieve a nearly constant relative phase delay of $\pi/2$ (or $3\pi/2$) over a broad spectral range while maintaining near-unity reflectivity for both orthogonal polarization components. We design these metasurface structures using full-wave electromagnetic simulations to target four distinct frequency windows of interest. The four devices are fabricated using standard photolithography and characterized using THz time-domain spectroscopy. The experimental results show excellent agreement with numerical simulations, which also allow further structural optimization through additional simulations. This work demonstrates robust broadband reflective THz quarter-wave plates, enabling advanced THz polarimetric applications and circularly polarized THz light--matter interactions in systems such as biomolecules, chiral phonons, and magnetic materials, as well as potential applications in THz communications and imaging. 

\section{\label{sec:design_simulations}Design and Simulations}
\subsection{\label{sec:principle}Metasurface Design Principle}
%\begin{figure}[t]
\begin{figure}[b]
  \centering
  \includegraphics[width=0.5\linewidth]{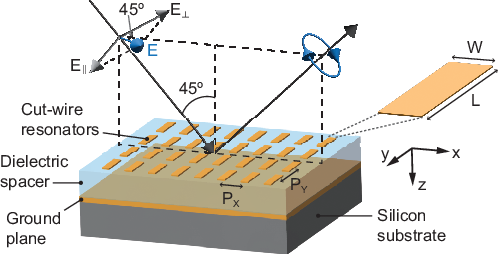}
  \caption{Schematic illustration of anisotropic metal-dielectric-metal metasurfaces functioning as broadband reflective quarter-wave plates operating in reflection. An incident THz wave linearly polarized at $45^\circ$ to the plane of incidence is converted into circularly polarized radiation over a broad spectral range.}
  \label{fig_devicescheme}
\end{figure}
\begin{figure}[ht]
  \centering
  \includegraphics[width=0.5\linewidth]{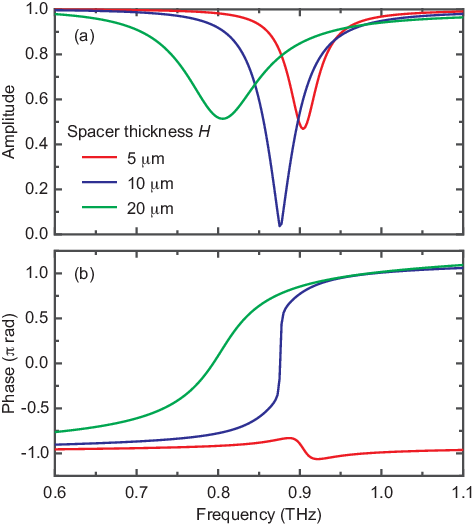}
  \caption{Simulated reflection amplitude (a) and phase (b) spectra of the first parallel-polarized Fabry-P\'{e}rot resonance of the metal-dielectric-metal metasurface for varying spacer thicknesses. }
  \label{fig_coupling}
\end{figure}

For clarity, we first describe the operating principle under normal incidence; in the experiments, however, the metasurfaces are designed for an incidence angle of $45^\circ$ to facilitate practical applications. The metasurface structure employed in this work is schematically illustrated in Fig.\,\ref{fig_devicescheme} and consists of a periodic array of gold cut-wire antennas separated from a gold ground plane by an SU-8 dielectric spacer. The reflection spectra, $\tilde{r}(\omega) = re^{i\phi}$, are obtained by numerical simulations in CST Studio Suite using unit cell boundary conditions, a perfect electrical conductor for the cut-wire antenna and ground plane, and a dielectric constant of 3.0 and loss tangent 0.05 at 1\,THz for the SU-8 spacer. The geometric parameters used are: $L = 100$ $\mu$m, $W = 10$ $\mu$m, and $P_x = P_y = 110$ $\mu$m. This anisotropic array-spacer-ground configuration forms a reflective Fabry-P\'{e}rot-like resonant cavity \cite{Chen_2012_OE}. Consequently, the reflection spectra exhibit a series of resonances that are different for the two orthogonal polarizations. For the parallel polarization that strongly interacts with the metasurface, the first Fabry-P\'{e}rot resonance is shown in Fig.\,\ref{fig_coupling} for different spacer thicknesses. The reflection amplitude spectra reveal a resonant dip [Fig.\,\ref{fig_coupling}(a)], accompanied by strong phase dispersion in the corresponding phase spectra [Fig.\,\ref{fig_coupling}(b)]. The spectral response, particularly the phase behavior, is governed by the interplay between intrinsic dissipative losses and radiative coupling to an external environment, both of which are strongly influenced by the spacer thickness. For a thin spacer, e.g., $H=5$\,$\mu$m, the cavity operates in the under-coupled regime, characterized by a shallow resonant reflection dip and a small Lorentzian-like phase variation across the resonance. As the spacer thickness increases, the resonant reflection dip deepens and eventually reaches zero at $H\approx10$\,$\mu$m, corresponding to the critical-coupling condition. With further increase of the spacer thickness, e.g., $H=20$\,$\mu$m, the cavity resonance transitions to the over-coupled regime: the reflection dip becomes shallower once again, while the phase undergoes a continuous $2\pi$ phase evolution across the resonance. 

We note that temporal coupled-mode theory (TCMT) is commonly employed to describe the resonant response in cavities, such as the array-spacer-ground metasurface structure used here, which can be effectively modeled as a single-port cavity \cite{Qu_2015_PRL}. TCMT is formally derived under the assumption of weak external coupling \cite{Haus_1984_Book}, a condition not strictly satisfied in our case, as indicated by the low quality factor and broad resonances. Nevertheless, the resonance behavior is well captured by TCMT expressions \cite{Chang_2019_PRL}, and we therefore adopt the standard terminology of under-coupling, critical-coupling, and over-coupling to describe the observed phenomena. 

%\begin{figure}[h!]
\begin{figure}[b]
  \centering
  \includegraphics[width=0.5\linewidth]{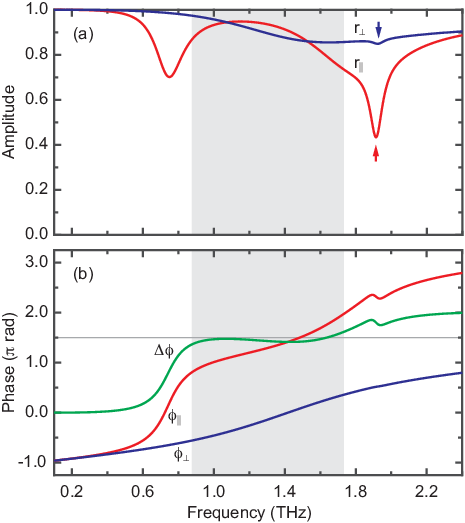}
  \caption{Simulated reflection amplitude (a) and phase (b) spectra of the reflective metasurface over an extended frequency range. Arrows indicate the resonant excitation of guided surface modes.}
  \label{fig_off_resonance}
\end{figure}
It is the over-coupled regime that we exploit to achieve broadband linear-to-circular polarization conversion. With spacer thickness of $H=30$ $\mu$m, in Fig.\,\ref{fig_off_resonance} we present the simulated reflection amplitude and phase responses over an extended frequency range covering the first few relevant Fabry-P\'{e}rot resonances. Both amplitude and phase spectra clearly exhibit pronounced birefringence for the parallel and perpendicular field components. For parallel polarization, two Fabry-P\'{e}rot resonances appear within the simulated frequency range, each operating in the over-coupled regime, as indicated by the continuous phase variation of approximately $2\pi$ across each resonance. The second Fabry-P\'{e}rot resonance manifests as a shoulder near 1.8\,THz, partially obscured by a sharp resonance at $1.9$\,THz (red arrow), which exhibits a deep reflection dip but minimal phase variation---likely resulting from resonant coupling to a surface wave mode \cite{Maier_2007_Plasmonics,Huang_2023_NatCommun}. For perpendicular polarization, only a single, much weaker and broader Fabry-P\'{e}rot resonance is observed within the simulated frequency range, accompanied by resonant excitation of a weaker surface wave mode (blue arrow) at a frequency similar to that of the parallel polarization.

Between the two Fabry-P\'{e}rot resonances of the parallel polarization, highlighted by the shaded region in Fig.\,\ref{fig_off_resonance}, the phase spectrum exhibits an approximately linear dispersion, with a slope closely matching that of the perpendicular polarization. Consequently, the relative phase delay, $\Delta \phi = \phi_{\parallel} - \phi_{\perp}$, remains nearly constant---$3\pi/2$ here---over a wide frequency range, where the reflection amplitudes are simultaneously high and nearly equal for both polarizations. As shown later, $\Delta\phi$ can be tuned to arbitrary values by tailoring the metasurface geometric parameters. This unique property is precisely what we exploit to enable broadband linear-to-circular polarization conversion in reflection.

\subsection{\label{sec:parametric}Parametric Study of Phase Dispersion}
%\begin{figure*}[t]
\begin{figure*}[b]
  \centering
  \includegraphics[width=0.9\linewidth]{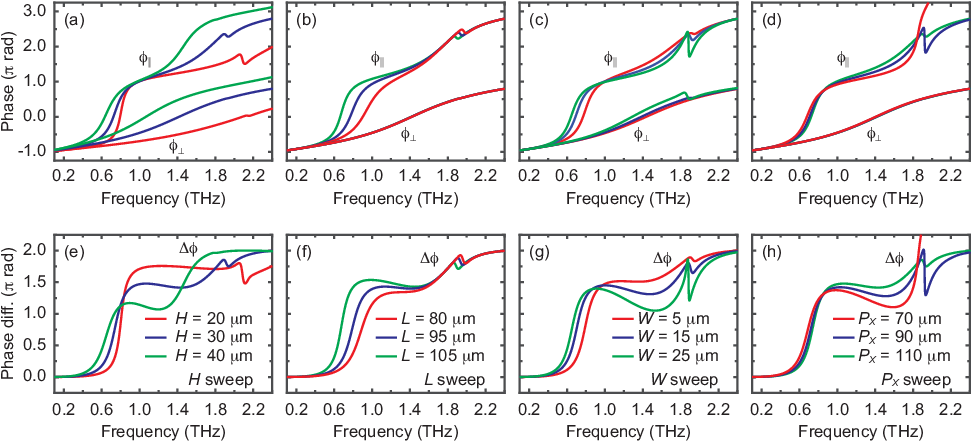}
  \caption{Simulated reflection phase spectra (a--d) and relative phase delay (e--h) by varying the geometric parameters: (a,e) spacer thickness $H$, (b,f) cut-wire length $L$, (c,g) cut-wire width $W$, and (d,h) lattice period in $x$-direction $P_x$.}
  \label{fig_parametric}
\end{figure*}
The geometric parameters available for tailoring the reflection amplitude and phase responses include the cut-wire length $L$ and width $W$, the lattice periods $P_x$ and $P_y$ along the \textit{x}- and \textit{y}-directions, respectively, and the spacer thickness $H$. Here, we fix the lattice period $P_y$ and investigate the effects of $H$, $L$, $W$ and $P_x$ on the reflection phase spectra. The goal is to achieve an approximately constant phase delay $\Delta\phi$ (e.g., $\pi/2$ or $3\pi/2$) between the two orthogonal polarizations over the widest possible bandwidth while maintaining their high reflection amplitudes. We first vary the spacer thickness $H$ while keeping all other parameters constant. The simulated reflection phase spectra and the corresponding relative phase delay are shown in Fig.\,\ref{fig_parametric}(a,e), respectively. It is evident that all relevant Fabry-P\'{e}rot resonances shift to lower frequencies as $H$ increases. In particular, the second resonance of the parallel polarization and the resonance of the perpendicular polarization exhibit a stronger dependence on $H$ compared with the first parallel-polarized resonance [Fig.\,\ref{fig_parametric}(a)]. This behavior can be understood as follows: The phase matching of the first Fabry-P\'{e}rot resonance is primarily determined by the resonant dispersion of the cut-wire array (which leads to ultrathin metamaterial perfect absorbers \cite{Chen_2012_OE}, for example), while the higher-order resonances require additional propagation phase, multiples of $2\pi$, provided solely by the spacer thickness. Consequently, varying $H$ mainly affects the value of $\Delta\phi$ and the achievable bandwidth, while having a relatively minor effect on its slope [Fig.\,\ref{fig_parametric}(e)].

Next, we vary the cut-wire length $L$, and the resulting reflection phase spectra and the corresponding relative phase delay are shown in Fig.\,\ref{fig_parametric}(b,f), respectively. As $L$ increases, the first parallel-polarized Fabry-P\'{e}rot resonance shifts to lower frequencies, accompanied by an increase in resonance strength, as suggested by the steeper phase variation across the resonance [Fig.\,\ref{fig_parametric}(b)]. In contrast, the phase responses of the second parallel-polarized resonance and the perpendicular-polarized resonance remain essentially unchanged. As a result, the spectral slope of $\phi_\parallel$ decreases with increasing $L$. The primary effect of varying $L$ is therefore to tune the flatness of $\Delta\phi$, resulting in an increase in $\Delta\phi$ on the low-frequency side, while leaving the high-frequency side largely unaffected [Fig.\,\ref{fig_parametric}(f)]. 

We then vary the cut-wire width $W$, and the resulting reflection phase spectra and corresponding relative phase delay are shown in Fig.\,\ref{fig_parametric}(c,g), respectively. The first parallel-polarized Fabry-P\'{e}rot resonance exhibits a slight redshift as $W$ increases, while the second parallel-polarized resonance and the perpendicular-polarized resonance remain largely unchanged [Fig.\,\ref{fig_parametric}(c)]. In contrast, increasing $W$ strengthens the resonant coupling to the surface modes for both polarizations, resulting in opposite changes in the phase spectral slopes within the frequency range of interest. Consequently, varying $W$ mainly modifies the spectral slope of $\Delta\phi$, with minimal effect on the low-frequency side but a more pronounced change on the high-frequency side [Fig.\,\ref{fig_parametric}(g)].  
 
Finally, we vary the lattice period $P_x$, which induces negligible changes in the first parallel-polarized and perpendicular-polarized Fabry-P\'{e}rot resonances, as shown in Fig.\,\ref{fig_parametric}(d,h) for the reflection phase spectra and corresponding relative phase delay, respectively. The perpendicular-polarized surface wave mode, although nearly invisible in $\phi_\perp$ and more discernible in $r_\perp$ (not shown), shifts favorably to higher frequencies as $P_x$ decreases. The second parallel-polarized Fabry-P\'{e}rot resonance appears to strengthen with decreasing $P_x$, likely arising from enhanced resonant coupling to a surface wave mode propagating along the $y$-direction, whose frequency is close. The increased coupling results from the higher density of cut-wire antennas at smaller $P_x$ [Fig.\,\ref{fig_parametric}(d)], while the surface wave mode along the $x$-direction is shifted to higher frequencies. Consequently, the spectral slope of $\phi_\parallel$---and thus $\Delta\phi$---decreases with decreasing $P_x$. $\Delta\phi$ exhibits negligible change on the low-frequency side, whereas the variation is more significant on the high-frequency side [Fig.,\ref{fig_parametric}(h)]. 

The off-resonance response for the parallel polarization, combined with the broad and relatively weak resonance for the perpendicular polarization, enables simultaneously high reflections for both orthogonal field components. The effects of the metasurface geometric parameters on the phase spectra are intricate and interdependent. While this complexity complicates the design process, it also provides substantial flexibility for achieving a broadband, near-constant relative phase delay. Although the preceding analysis assumes normal incidence, in the following we further tailor the geometric parameters to achieve high reflection and a near-constant relative phase delay over a broad spectral range at an oblique incidence angle, e.g., $45^\circ$, which is more relevant for practical applications.

\subsection{\label{sec:simulations}Metasurface Design Under $45^\circ$ Incidence}
Under an incidence angle of $45^\circ$, the reflective metasurface can be used like a mirror while functioning as a quarter-wave plate. Owing to its anisotropic geometry, two orientations of the plane of incidence are possible, with the cut-wire either parallel or perpendicular to the incident plane. The latter configuration---schematically illustrated in Fig.\,\ref{fig_devicescheme}---turns out to provide a broader operational bandwidth and is therefore adopted in the following analysis. In this configuration, the incident electric field is oriented at $\pm45^\circ$ with respect to the plane of incidence, yielding field components parallel and perpendicular to the cut-wires. In numerical simulations, we systematically vary the geometric parameters to tune both amplitude and phase responses, aiming to achieve a near-constant relative phase delay $\Delta\phi$ while maintaining high reflection amplitudes. We first present a representative design covering a specific portion of the THz spectrum, and the remaining spectral regions are subsequently addressed by further parameter tuning.

\begin{figure}[b]
  \centering
  \includegraphics[width=0.5\linewidth]{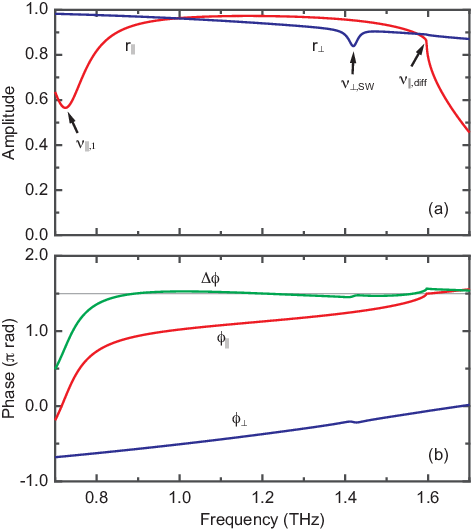}
  \caption{Simulated reflection amplitude (a) and phase (b) spectra of a metasurface designed under $45^\circ$ incidence.}
  \label{fig_device_3_ampl_phase}
\end{figure}
For simplicity, we adopt a square lattice (i.e., $P_x = P_y = P$). Using the parameters $P=110$ $\mu$m, $L=100$ $\mu$m, $W=15$ $\mu$m, and $H=27$ $\mu$m, the simulated reflection amplitude and phase spectra are shown in Fig.\,\ref{fig_device_3_ampl_phase}. Within the simulated frequency range, the parallel polarization exhibits its first Fabry-P\'{e}rot resonance at $\nu_{\parallel,1}=0.72$\,THz. At higher frequencies, coupling to high-order diffraction modes (i.e., diffraction order $|m| \ge 1$) begins to occur at $\nu_{\parallel,\mathrm{diff}}=1.60$\,THz, preceding the second Fabry-P\'{e}rot resonance. For the perpendicular polarization, weak resonant coupling to a surface wave mode occurs at $\nu_{\perp,\mathrm{SW}}=1.42$\,THz, which appears as a small reflection dip in Fig.\,\ref{fig_device_3_ampl_phase}(a). Both the high-order diffraction modes and the surface wave mode have negligible effect on the overall phase dispersion. In the frequency range between 0.82\,THz and 1.7\,THz, the phase spectra of the two orthogonal polarizations exhibit an approximately linear dependence on frequency, as shown in Fig.\,\ref{fig_device_3_ampl_phase}(b). The two phase curves are nearly parallel, yielding a near-constant $\Delta\phi = 3\pi/2$. Meanwhile, the reflection amplitudes remain high throughout most of this range, except near frequencies where diffraction occurs. 

\begin{figure}[b]
  \centering
  \includegraphics[width=0.5\linewidth]{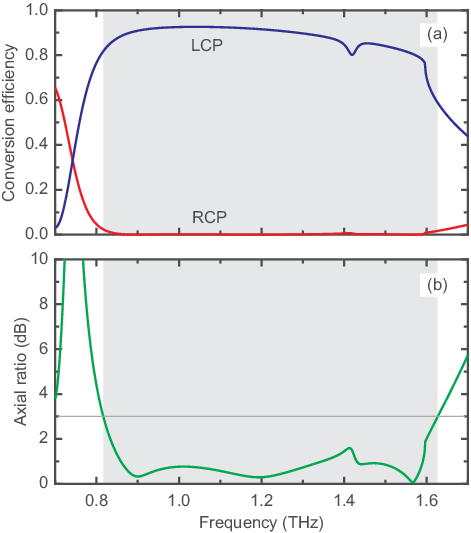}
  \caption{(a) Conversion efficiencies to left- and right-handed circular polarizations, retrieved from the linear polarization reflection coefficients in Fig.\,\ref{fig_device_3_ampl_phase}. (b) The corresponding calculated axial ratio on a logarithmic scale. }
  \label{fig_device_3_circular_polarization}
\end{figure}
When the incident electric field is linearly polarized at $+45^\circ$ with respect to the plane of incidence, the reflected field can be expressed in the circular polarization basis. In terms of the reflection coefficients for the parallel and perpendicular components, $\tilde{r}_\parallel = r_\parallel e^{i\phi_\parallel}$ and $\tilde{r}_\perp = r_\perp e^{i\phi_\perp}$, the left-handed circular polarization (LCP) and right-handed circular polarization (RCP) reflection coefficients are given by 
\begin{eqnarray}
    \tilde{r}_{\mathrm{LCP}} = \frac{\tilde{r}_{\parallel}-i\tilde{r}_{\perp}}{2}, \label{eq:LCP_from_LP}\\
    \tilde{r}_{\mathrm{RCP}} = \frac{\tilde{r}_{\parallel}+i\tilde{r}_{\perp}}{2}. \label{eq:RCP_from_LP}
\end{eqnarray}
The corresponding circularly polarized power reflection coefficients are defined as $R_{\mathrm{LCP}} = |\tilde{r}_{\mathrm{LCP}}|^2$ and $R_{\mathrm{RCP}} = |\tilde{r}_{\mathrm{RCP}}|^2$. The degree of circular polarization is quantified by the axial ratio (AR), which is defined on a logarithmic scale by 
\begin{equation}
    AR = 20\log{\left(\frac{a}{b}\right)}, \label{eq:AR}
\end{equation}
where $a$ and $b$ represent the lengths of the major and minor axes of the reflected polarization ellipse, respectively, and are given by
\begin{eqnarray}
    a^2 = \frac{1}{2}\left[r_{\parallel}^{2}+r_{\perp}^{2}+\sqrt{r_{\parallel}^{4}+r_{\perp}^{4}+2r_{\parallel}^{2}r_{\perp}^{2}\cos\!\left(2\Delta\phi\right)}\right],\\
    b^2 = \frac{1}{2}\left[r_{\parallel}^{2}+r_{\perp}^{2}-\sqrt{r_{\parallel}^{4}+r_{\perp}^{4}+2r_{\parallel}^{2}r_{\perp}^{2}\cos\!\left(2\Delta\phi\right)}\right].
    \label{eq:ab}
\end{eqnarray}

The retrieved circularly polarized power reflection coefficients (i.e., conversion efficiencies) are presented in Fig.\,\ref{fig_device_3_circular_polarization}(a). The results show that the incident THz waves linearly polarized at $+45^\circ$ are efficiently converted to left-handed circular polarization with a conversion efficiency of 80--90\% over a broad frequency range from approximately 0.8 to 1.6\,THz, while the right-handed circular polarization component remains negligible within this frequency range. A wave is generally considered circularly polarized when $AR \le 3$\,dB. As shown in Fig.\,\ref{fig_device_3_circular_polarization}(b), the corresponding calculated axial ratio stays below 3\,dB---and is mostly below 1\,dB---over an octave-spanning bandwidth from 0.82 to 1.63\,THz. This confirms that the reflected THz waves meet the circular polarization criterion and demonstrate high polarization purity of the converted THz waves. Note that when the incident polarization is rotated by $90^\circ$ (i.e., $-45^\circ$ with respect to the plane of incidence), the handedness of the converted waves reverses, and the reflected THz radiation becomes right-handed circularly polarized. This property enables circular polarization modulation, for instance, by electrically switching the incident THz linear polarization between $\pm45^\circ$ directions \cite{Hirota_2006_OE}.

\section{Fabrication Procedures}
The metasurfaces are fabricated on test-grade 4-inch silicon wafers. The design parameters for four complementary devices are summarized in Table~\ref{fabparameter}. First, a 5\,nm titanium adhesion layer and a 200\,nm gold ground plane are deposited on the silicon wafer using electron-beam evaporation. Next, SU-8 photoresist is spin-coated to the target thickness ($H$) by adjusting the spin speed. The SU-8 layer is hard-baked by ramping the temperature from 60\,$^\circ$C to 200\,$^\circ$C over \textasciitilde45~min, holding at 200\,$^\circ$C for \textasciitilde30\,min, and then cooling back to 60\,$^\circ$C over \textasciitilde45~min to make it mechanically robust and chemically stable.
%\begin{table}[b]
\begin{table}[ht]
  \caption{\label{fabparameter}Design parameters of broadband THz linear-to-circular polarization converters. $P_x$ and $P_y$ are the lattice periods along the $x$- and $y$-directions, respectively, $L$ and $W$ are the cut-wire length and width, respectively, and $H$ is the spacer thickness. $\nu_\mathrm{op}$ represents designed operational frequency range.} 
  \begin{ruledtabular}
  \begin{tabular}{ccccc}
  \mbox{ }&\mbox{Device 1}&\mbox{Device 2}&\mbox{Device 3}&\mbox{Device 4}\\
  \hline
  $P_x~\&~P_y$ ($\mu$m)&350&195&110&55\\
  $L$ ($\mu$m)&340& 185 & 100 & 50\\
  $W$ ($\mu$m)&15& 15 & 15 & 7\\
  $H$ ($\mu$m)&100& 53 & 27 & 13\\
  $\nu_\mathrm{op}$ (THz)&0.25--0.47&0.45--0.88 &0.82--1.63 &1.65--3.29\\
  \end{tabular}
  \end{ruledtabular}
\end{table}

To fabricate the top cut-wire antenna array, we prepare a bilayer lift-off stack by spin-coating LOR-10B followed by AZ5214 onto the cured SU-8 film. The metasurface pattern is defined by ultraviolet exposure through a predesigned chrome mask and subsequent development. This process creates clean openings down to the cured SU-8 and an undercut profile suitable for the lift-off process. Next, the top metal layer is deposited by electron-beam evaporation of 5\,nm titanium and 200\,nm gold. Lift-off is carried out by soaking the samples in acetone overnight, leaving anisotropic gold cut-wire antennas patterned on the SU-8 dielectric spacer above the continuous gold ground plane. Any residual photoresist is removed by rinsing the samples in Remover~PG for approximately 5\,min.

\section{Spectral Characterizations}
%\begin{figure}[ht]
\begin{figure}[b]
  \centering
  \includegraphics[width=0.5\linewidth]{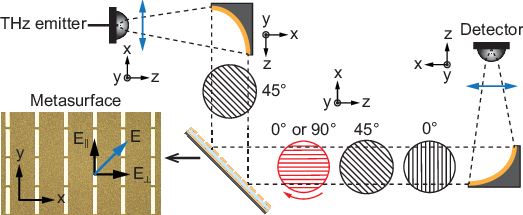}
  \caption{Schematic of the THz time-domain spectroscopy system for characterizing the linear-to-circular polarization conversion performance of the reflective metasurfaces.}
  \label{fig_Schematic}
\end{figure}
The fabricated metasurfaces are characterized using a home-built, all-fiber-coupled THz time-domain spectroscopy system schematically illustrated in Fig.\,\ref{fig_Schematic}. Driven by telecom femtosecond laser pulses with wavelengths centered at 1560\,nm, the photoconductive THz emitter generates THz pulses linearly polarized along the $x$-direction (parallel to the optical table), while the second photoconductive antenna serves as the THz detector and is configured to probe the same $x$-polarized THz radiation. The metasurfaces are mounted such that their cut-wire major axis is aligned along the $y$-direction (perpendicular to the optical table), which defines the $E_{\parallel}$ component. The orthogonal field component is correspondingly defined as $E_{\perp}$. An off-axis parabolic mirror collimates the emitted THz beam and directs it onto the metasurface at an incidence angle of $45^\circ$. A wire-grid polarizer placed before the metasurface is oriented at either $+45^{\circ}$ or $-45^{\circ}$ with respect to the plane of incidence, thereby producing linearly polarized incident THz waves with equal-amplitude components $E_{\parallel}$ and $E_{\perp}$. The specularly reflected THz beam is analyzed using three wire-grid polarizers. The first analyzer is oriented at either $90^{\circ}$ or $0^{\circ}$ to select the reflected $E_{\parallel}$ or $E_{\perp}$ component, respectively. The orientations of the second and third analyzers are fixed at $45^\circ$ and $0^\circ$, respectively. Together, these analyzers project the selected field component onto the detector's polarization axis, enabling independent measurements of $E_{\parallel}$ and $E_{\perp}$ without requiring rotation of the detector.

%\begin{figure*}%[t]
\begin{figure*}[b]
  \centering
  \includegraphics[width=0.9\linewidth]{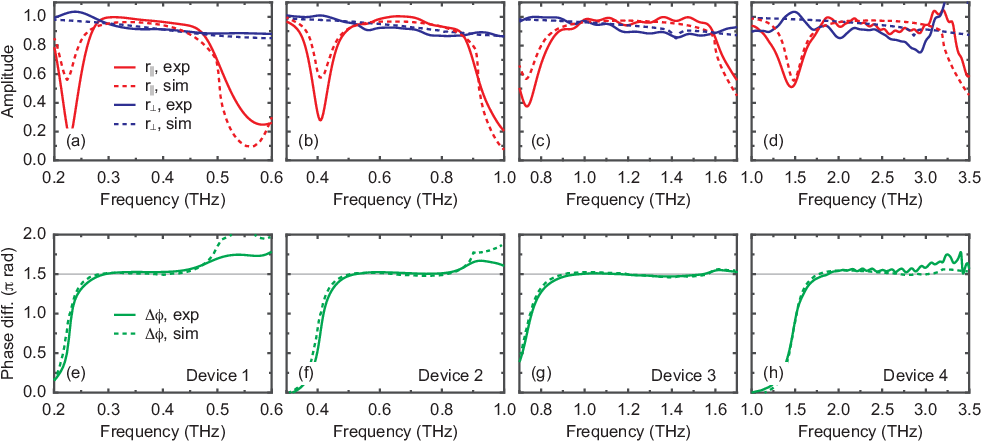}
\caption{Experimentally measured (solid curves) and numerically simulated (dashed curves) reflection amplitude (a--d) and relative phase delay (e--h) for metasurface devices 1--4. Panels (a,e), (b,f), (c,g) and (d,h) correspond to devices 1, 2, 3, and 4, respectively.}
\label{fig_RandPD}
\end{figure*}
The experimentally measured reflection amplitude spectra for the parallel and perpendicular field components are presented in Fig.\,\ref{fig_RandPD}(a--d) for devices 1--4, respectively. Numerically simulated results are also included as dashed curves for comparison. For each metasurface device, both the experimental and simulated results show that the parallel-polarized reflection exhibits an over-coupled Fabry-P\'{e}rot cavity resonance at $\nu_{\parallel,1}$ on the low-frequency side. On the high-frequency side, coupling to high-order diffraction modes sets in at $\nu_{\parallel,\mathrm{diff}}$, as evidenced by the abrupt drop in $r_\parallel$ above $\nu_{\parallel,\mathrm{diff}}$. Specifically, $\nu_{\parallel,\mathrm{diff}} = 0.50$\,THz for device 1, 0.90\,THz for device 2, 1.59\,THz for device 3, and 3.20\,THz for device 4. Between $\nu_{\parallel,1}$ and $\nu_{\parallel,\mathrm{diff}}$, there is a broad frequency range where the reflection amplitude $r_\parallel$ remains high. The results also show that $r_{\perp}$ stays close to unity across the measured and simulated spectral range and does not exhibit significant resonance behavior, as its corresponding resonance occurs at a much higher frequency beyond the simulated frequency range. The corresponding relative phase delay $\Delta\phi$ is shown in Fig.\,\ref{fig_RandPD}(e--h). All four devices exhibit a nearly constant phase delay of $3\pi/2$ (equivalently, $-\pi/2$) over a broad frequency range beyond the first Fabry-P\'{e}rot resonance, whereas the second Fabry-P\'{e}rot resonance is effectively obscured by the coupling to high-order diffraction modes. These observations thereby verify the essential conditions for high-efficiency, broadband THz linear-to-circular polarization conversion: both orthogonal THz field components exhibit high reflection amplitudes over a wide spectral range while maintaining a nearly constant relative phase delay of $\pm\pi/2$.
%\begin{figure*}%[h!]
\begin{figure*}[b]
  \centering
  \includegraphics[width=0.9\linewidth]{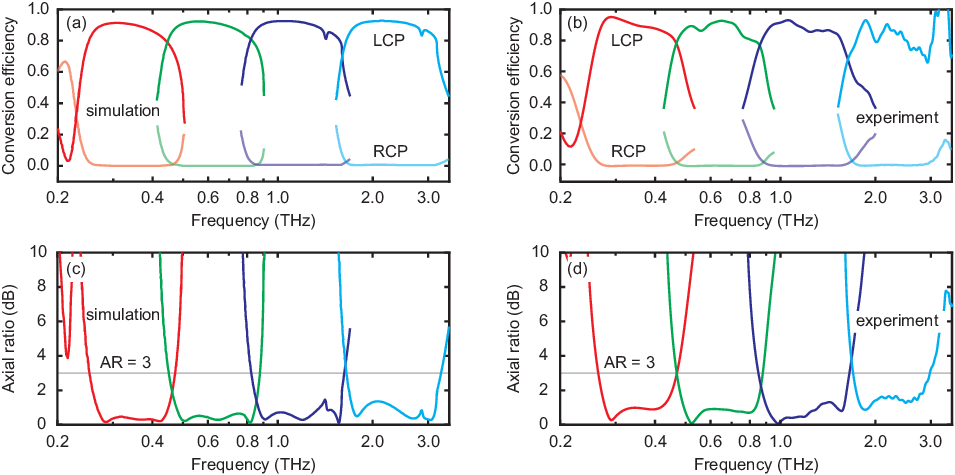}
  \caption{Performance assessment of linear-to-circular polarization conversion for the four metasurface devices. (a) Simulated and (b) measured conversion efficiencies for devices 1--4, indicated by red, green, blue, and cyan curves, respectively. (c) Simulated and (d) measured axial ratio, quantifying the circular polarization purity across the operational bandwidth.}
  \label{fig_CEandAR}
\end{figure*}

Since typical THz time-domain spectroscopy systems cannot directly resolve circular polarization states, the circularly polarized reflection components are reconstructed from the measured linearly polarized amplitude and phase spectra using Eqs.~(\ref{eq:LCP_from_LP}) and (\ref{eq:RCP_from_LP}). The resulting power conversion efficiencies are presented in Fig.~\ref{fig_CEandAR}(a) and (b) for numerical simulations and experimental measurements, respectively. The results clearly show that an incident wave linearly polarized at $+45^\circ$ with respect to the plane of incidence is efficiently converted into left-handed circular polarization over a broad spectral range within the bandwidth of the incident THz radiation. The four complementary metasurfaces collectively span the entire 0.25--3\,THz frequency range accessible to a typical THz time-domain spectrometer. A conversion efficiency exceeding 80\% is achieved across most of this spectral range, with peak efficiencies exceeding 90\% and even 95\% at some selected frequencies, highlighting the broadband, high-efficiency performance of the fabricated devices.

The negligible conversion to the opposite circular polarization, as shown in Fig.~\ref{fig_CEandAR}(a) and (b), indicates the high purity of the generated circular polarization of the output THz waves within the frequency ranges of interest. The axial ratio, defined in Eq.\,(\ref{eq:AR}), is commonly used to quantify the purity of circular polarization. The calculated axial ratios for the four metasurface devices are presented in Fig.\,\ref{fig_CEandAR}(c) and (d) for numerical simulations and experimental measurements, respectively, thereby providing a quantitative assessment of the circular polarization performance across the broadband THz spectral range. The results show that, over most of the targeted spectral regions, the generated circular polarization exhibits $AR<1$\,dB, indicating very high circular polarization purity. Using the standard criterion of circular polarization, $AR \leq 3$\,dB, the experimentally measured bandwidths of the four metasurfaces are 0.26--0.47\,THz, 0.46--0.88\,THz, 0.85--1.65\,THz, and 1.69--3.00\,THz, respectively. The corresponding simulated bandwidths are 0.25--0.47\,THz, 0.45--0.88\,THz, 0.82--1.62\,THz, and 1.65--3.29\,THz, respectively. Each device therefore achieves an approximately octave-spanning bandwidth of circular polarization, and together the four devices cover the entire 0.25--3\,THz spectral range with near-unity conversion efficiency. 

\section{Discussion}
The excellent quantitative agreement between experiment and simulation for all four devices enables further optimization of metasurface designs through numerical modeling for an even broader spectral response. In the preceding structures, the primary bandwidth limitations arise from the resonant coupling to surface waves and the onset of high-order diffraction, as well as the pursuit of maintaining a low axial ratio for high-purity circular polarization. At an incidence angle of $45^\circ$, diffraction begins at lower frequencies than under normal incidence, thereby constraining the usable bandwidth. To suppress both diffraction and surface waves, an effective approach is to reduce the lattice period $P_x$, which shifts the onset of diffraction to higher frequencies. This strategy can also affect the resonant coupling to surface waves, which are known to be sensitive to the angle of incidence \cite{Maier_2007_Plasmonics}. 

\begin{figure}[t]
  \centering
  \includegraphics[width=0.5\linewidth]{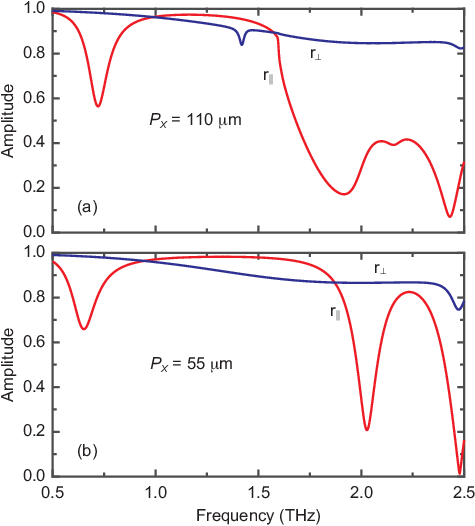}
  \caption{Simulated reflection amplitude spectra for lattice period (a) $P_x = 110$\,$\mu$m and (b) $P_x = 55$\,$\mu$m, illustrating the shift of the diffraction onset and surface wave resonances to higher frequencies.}
  \label{fig_bandwidth_px}
\end{figure}
Simulation results validate this approach, as shown in Fig.\,\ref{fig_bandwidth_px}. By reducing the lattice period $P_x$ while keeping the other geometric parameters identical to those in Fig.\,\ref{fig_device_3_ampl_phase}, the onset of parallel-polarized coupling to high-order diffraction, originally at $\nu_{\parallel,\mathrm{diff}} = 1.6$\,THz for $P_x = 110$\,$\mu$m [Fig.\,\ref{fig_bandwidth_px}(a)], shifts to higher frequencies beyond the simulated spectral range when $P_x = 55$\,$\mu$m {Fig.\,\ref{fig_bandwidth_px}(b)}. Simultaneously, the perpendicular-polarized resonant coupling to the surface wave mode shifts from $\nu_{\perp,\mathrm{SW}} = 1.42$\,THz for $P_x = 110$\,$\mu$m to 2.48\,THz for $P_x = 55$\,$\mu$m. Because the surface wave mode occurs at the same frequency for both polarizations, the strong resonance at 2.48\,THz for parallel polarization is attributed to resonant surface-wave coupling, whereas the resonance at 2.03\,THz corresponds to the second Fabry-P\'{e}rot resonance [Fig.,\ref{fig_bandwidth_px}(b)].
%\begin{figure}%[b]
\begin{figure}
  \centering
  \includegraphics[width=0.5\linewidth]{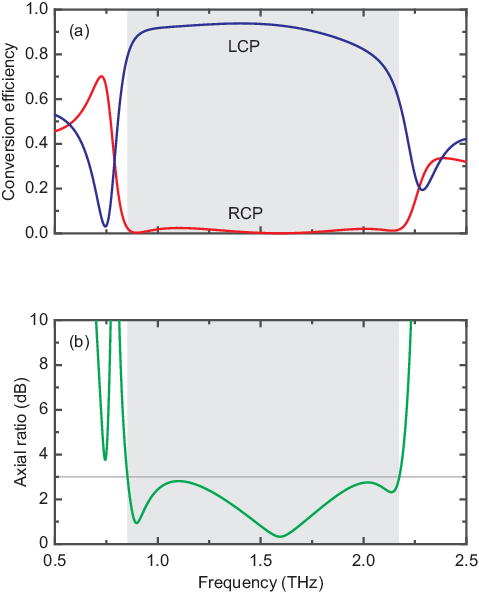}
  \caption{(a) Simulated linear-to-circular polarization conversion efficiencies and (b) corresponding axial ratio for the optimized metasurface design. }
  \label{fig_broader_optimized}
\end{figure}

Through systematic optimization of $P_x$ and related structural parameters, it is possible to extend the high-reflection region while maintaining a near-constant relative phase delay between the two orthogonal field components, thereby enhancing both the bandwidth and efficiency of linear-to-circular polarization conversion. Considering a 3-dB axial-ratio tolerance and employing iterative numerical optimization, we arrive at an optimized metasurface design with $P_x = 55$\,$\mu$m, $P_y = 110$\,$\mu$m, $L = 98$\,$\mu$m, $W = 12$\,$\mu$m, and $H = 19$\,$\mu$m. The retrieved power conversion efficiencies for left- and right-handed circular polarizations are presented in Fig.\,\ref{fig_broader_optimized}(a), while the corresponding axial ratio is shown in Fig.\,\ref{fig_broader_optimized}(b). The optimized design exhibits a 3-dB axial-ratio bandwidth of 0.85--2.17\,THz, significantly exceeding one octave and demonstrating substantially improved broadband circular polarization performance. Additional simulations (not shown) reveal that the bandwidth can be further improved by reducing the dielectric constant of the spacer material. 

We note that the trial-and-error optimization process is inherently tedious and computationally demanding, as the effects of the geometric parameters are highly interdependent and nonlinearly coupled. Variations in one parameter can significantly affect the impact of others, making it difficult to intuitively predict the overall spectral response. Consequently, achieving an optimal design requires extensive parametric sweeps and iterative refinement. In this regard, rapidly advancing artificial intelligence techniques—particularly deep learning with deep neural networks—offer a promising pathway for metasurface design and performance optimization \cite{Nadell_2019_OE}. By learning the complex relationship between structural parameters and electromagnetic responses, such data-driven methods can greatly accelerate the inverse design process and reduce the computational burden compared to conventional brute-force strategies.

\section{Conclusion}
In conclusion, we have demonstrated high-performance broadband THz quarter-wave plates based on over-coupled metal-dielectric-metal reflective metasurfaces. By harnessing the phase dispersion associated with anisotropic over-coupled Fabry-P\'{e}rot cavity resonances and precisely engineering the relative phase delay between orthogonal field components, we achieve efficient linear-to-circular polarization conversion over a broad THz spectral range. Four complementary metasurface designs collectively span the 0.25--3\,THz frequency range accessible to a typical THz time-domain spectroscopy system, with each device providing an approximately octave-wide bandwidth characterized by an axial ratio below 3\,dB and conversion efficiencies exceeding 80\% across most of the operating band. The metasurface structures presented here are compatible with straightforward wafer-scale fabrication and exhibit excellent tolerance to geometric deviations arising during the fabrication process, underscoring their practical viability. The excellent agreement between experimental measurements and numerical simulations validates the robustness and reliability of our design strategy and further enables systematic optimization of structural parameters toward enhanced device performance. In particular, reducing the lattice period effectively suppresses coupling to diffraction and surface waves under oblique incidence, thereby extending the usable bandwidth while preserving the required phase relationship. These results establish over-coupled reflective metasurfaces as a versatile and powerful platform for broadband THz polarization control. Importantly, the nearly constant phase delay enabled by this approach can, in principle,  be engineered to any target value between 0 and $2\pi$. The demonstrated approach provides a scalable pathway toward compact, efficient, and broadband THz polarization components for spectroscopy, imaging, communications, and diverse light--matter interaction studies.

\begin{acknowledgments}
 We acknowledge partial support provided by the Los Alamos National Laboratory Laboratory Directed Research and Development (LDRD) program. This work was primarily performed at the Center for Integrated Nanotechnologies, an Office of Science User Facility operated for the U.S. DOE Office of Science. Los Alamos National Laboratory, an affirmative action equal opportunity employer, is managed by Triad National Security, LLC for the US DOE NNSA, under contract no. 89233218CNA000001.
\end{acknowledgments}

\section*{Data Availability Statement}
The data that support the findings of this study are available from the corresponding author upon reasonable request.

%\bibliography{ChenLANL}% Produces the bibliography via BibTeX.

\end{document}